\documentstyle[epsfig,12pt]{article}

\newcommand{\be}{\begin{equation}}
\newcommand{\ee}{\end{equation}}
\newcommand{\ba}{\begin{eqnarray}}
\newcommand{\ea}{\end{eqnarray}}
\newcommand{\bd}{\begin{displaymath}}
\newcommand{\ed}{\end{displaymath}}

\def\thalf{{\textstyle{\frac{1}{2}}}}

\def\oneth{{\textstyle{\frac{1}{3}}}}
\def\twoth{{\textstyle{\frac{2}{3}}}}

\def\fourth{{\textstyle{\frac{4}{3}}}}

\title{{\bf Viscous Properties of Strongly Interacting Matter at High Temperature}}
\author{{Joseph I. Kapusta} \vspace*{0.1in}\\
{\it School of Physics and Astronomy, University of Minnesota}\\
 {\it Minneapolis, Minnesota 55455, USA}}

\date{December 9, 2008}

\begin{document}

\maketitle

\begin{abstract}

Substantial collective flow is observed in collisions between large nuclei at 
high energy, and the data are well-reproduced by perfect fluid dynamics.  In a separate development, calculation of the dimensionless ratio of shear viscosity $\eta$ to entropy density within AdS/CFT yields $\eta/s = 1/4\pi$, which has been conjectured to be a lower bound for any physical system.  It is shown that the transition from hadrons to quarks and gluons has behavior similar to helium, nitrogen, and water at and near their phase transitions in the ratio $\eta/s$.  Conversely, there are indications that the ratio of bulk viscosity $\zeta$ to entropy density may have a maximum in the vicinity of the phase transition. Therefore it is possible that experimental measurements can pinpoint the location of the transition or rapid crossover in QCD via the ratios $\eta/s$ and $\zeta/s$ in addition to and independently of the equation of state.   

\end{abstract}

\vspace*{0.5in}

\section{Introduction}

A transition from a phase of hadrons to a phase of quarks and gluons with increasing temperature and/or baryon chemical potential has been studied theoretically for three decades \cite{KMR}.  Calculations with effective hadronic field theories and with perturbative QCD give consistent and intuitively reasonable estimates of where this transition occurs, but at the present time only lattice gauge theory calculations on lattices of finite size can yield quantitatively accurate numbers.  For two flavors of massless quarks the phase transition is second order.  For three flavors of massless quarks the phase transition is first order.  For the real world with nonzero masses for up, down and strange quarks the answer is not definitively known yet, but the answer is likely to be a rapid crossover from one phase to another without a rigorous thermodynamic phase transition, at least for zero baryon chemical potential \cite{latticeQCD}.  Indeed, there may be a line of first order phase transition of the plane of temperature $T$ versus baryon chemical potential $\mu$ starting from the chemical potential axis and terminating at some critical point in the $T - \mu$ plane \cite{fodor02,Forcrand,ejiri03}.  The Relativistic Heavy Ion Collider (RHIC) at Brookhaven National Laboratory was constructed explicitly to create quark gluon plasma.  After nine years of operation, what have the experiments told us?    

One of the amazing discoveries of experimental measurements of gold on gold 
collisions at RHIC is the surprising amount of collective flow exhibited by the outgoing hadrons.  Collective flow is evidenced in both the single-particle transverse momentum distribution \cite{radial}, commonly referred to as radial flow, and in the asymmetric azimuthal distribution around the beam axis \cite{RHICv2}, commonly referred to as elliptic flow and characterized by a Fourier coefficient called $v_2$. Elliptic flow was expected to be much smaller at RHIC than it was at the lower energies of the SPS (Super Proton Synchrotron) at CERN \cite{SPSv2}; in fact, it is about twice as large.  Various theoretical calculations \cite{coalesce} support the notion that collective flow is mostly generated early in the nucleus-nucleus collision and is present before partons coalesce or fragment into hadrons.  Theoretical calculations including only 
two-body interactions between partons cannot generate sufficient flow to explain the observations unless partonic cross sections are artificially enhanced by more than an order of magnitude over perturbative QCD predictions \cite{Molnar}.  Thus quark-gluon matter created in these collisions is strongly interacting, unlike the type of weakly interacting quark-gluon plasma expected to occur at very high temperatures on the basis of asymptotic freedom \cite{pQCD}.  On the other hand, lattice QCD calculations yield an equation of state that differs from an ideal gas only by about 10\% once the temperature exceeds $1.5 T_c$, where $T_c \approx 175$ MeV is the critical or crossover temperature from quarks and gluons to hadrons \cite{latticeQCD}.  Furthermore, perfect fluid dynamics with zero shear and bulk viscosities reproduces the measurements of radial flow and $v_2$ quite well up to transverse momenta of order 1.5 GeV/c \cite{hydro}.  Parametric fits to the transverse momentum spectra of hadrons, such as pions, kaons, and protons, result in average transverse fluid flow velocities of more than 50\% the speed of light and chemical freezeout temperatures on the order of 160 MeV.  These results have been interpreted as strong indicators of early thermalization and collective flow on a time scale of several fm/c.

An amazing theoretical discovery was made by Kovtun, Son and Starinets \cite{Son}.  They showed that certain special field theories, special in the sense that they are dual to black branes in higher space-time dimensions, have the ratio $\eta/s = 1/4\pi$ (in units with $\hbar = k_{\rm B} = c =1$) where $\eta$ is the shear viscosity and $s$ is the entropy density.  
The connection between transport coefficients and gravity is intuitively clear since both involve (commutators of) the stress-energy-momentum tensor $T^{ij}$.
They conjectured that all substances have this value as a lower limit, and gave as examples helium, nitrogen, and water at pressures of 0.1 MPa, 10 MPa, and 100 MPa, respectively.  Is the RHIC data, represented by radial and elliptic flow, telling us that the created matter has a very small viscosity, that it is a 
{\it perfect fluid}?

The relatively good agreement between perfect fluid calculations and experimental data for hadrons of low to medium transverse momentum at RHIC suggests that the viscosity is small; however, it cannot be zero.  Indeed, the calculations within AdS/CFT suggests that $\eta \ge s/4\pi$.  It will now be argued that sufficiently precise calculations and measurements should allow for a determination of the ratio $\eta/s$ as well as the ratio of bulk viscosity to entropy density $\zeta/s$ as functions of temperature, and that these ratios can pinpoint the location of the phase transition or rapid crossover from hadronic to quark and gluon matter.  This is a different method than trying to infer the equation of state of QCD in the form of pressure $P$ as a function of temperature $T$ or energy density $\epsilon$.

The shear and bulk viscosities are rigorously given by the Kubo formulas which are based on linear response theory \cite{Kubo}.
\ba
\eta &=& \frac{1}{20} \lim_{\omega \rightarrow 0}
\frac{1}{\omega} \int d^4x e^{i\omega t} \langle
\left[ T^{ij}_{\rm traceless}(x), T^{ij}_{\rm traceless}(0) \right] \rangle
\theta(t) \\
\zeta &=& \frac{1}{2} \lim_{\omega \rightarrow 0} \frac{1}{\omega}
\int d^4x \, {\rm e}^{i\omega t}
\langle \left[ {\cal P}(t,{\bf x}), {\cal P}(0,{\bf 0})
\right] \rangle \theta(t) 
\ea
Here $T^{ij}_{\rm traceless}$ represents the traceless part of the spatial components of the energy-momentum tensor and ${\cal P} = \oneth T^i_{\;\; i}$ represents the trace of the tensor (the pressure in equilibrium).  There are some interesting limiting cases.

Why is the entropy per baryon of the universe as large as $10^9$?  It had originally been presumed that the expansion of the universe was adiabatic.  Although shear viscosity and thermal conductivity play no role in a 
Robertson-Walker model, bulk viscosity could.  In order to address this problem, as well as the problem of damping of protogalactic fluctuations immediately prior to the period of recombination of hydrogen, Weinberg was led to study the bulk viscosity due to radiation \cite{Weinberg1971}.  Much earlier Thomas \cite{Thomas} had calculated the correction to the energy-momentum tensor due to radiation, meaning photons, neutrinos, or gravitons, interacting with material particles with much smaller mean free times.  From that result Weinberg inferred the shear and bulk viscosities.  (He also inferred the heat conductivity which is of less interest for RHIC or LHC, the Large Hadron Collider at CERN.) For photons
\ba
\eta &=& \frac{4}{15} a T^4 \tau_{\gamma} \label{Weinbergshear} \\
\zeta &=& 4 a T^4 \tau_{\gamma} \left( \oneth - (\partial P/\partial \epsilon)_n \right)^2
\label{Weinbergbulk}
\ea
where $a = \pi^2/15$, $\tau_{\gamma}$ is the photon mean free time, and the derivative $(\partial P/\partial \epsilon)_n$ is taken at constant baryon density.  For neutrinos these results are multiplied by a factor of 7/8.  This leads to the oft-quoted result $\zeta = 15(\oneth - v_s^2)^2 \eta$.  If the material particles themselves are nearly massless then $\zeta/\eta \rightarrow 0$.  One must always keep in mind the conditions for which these results apply, namely, massless particles with a very large mean free time compared to material particles.  In heavy ion collisions the photons escape from the system without interacting, so this formula is not directly relevant for RHIC and LHC. 

Gavin \cite{Gavin1985} derived relativistic formulas for a single species of hadrons in the relaxation time approximation.  
\ba
\eta &=& \frac{\tau}{15T} \int \frac{d^3p}{(2\pi)^3} \frac{|{\bf p}|^4}{E^2} f^{\rm eq}(E/T) \label{seanshear} \\
\zeta &=& \frac{\tau}{9T} \int \frac{d^3p}{(2\pi)^3} \frac{1}{E^2}
\left[ \left( 1 - 3v_s^2 \right) E^2 - m^2 \right]^2
f^{\rm eq}(E/T) \, .
\label{seanbulk}
\ea
The factor $(\oneth - v_s^2)^2$ appears once again, but here $v_s$ refers to the speed of sound of the single species of particles themselves.  There is no additional radiation that provides the transport.  The bulk viscosity still has the property that it goes to zero as $v_s^2 \rightarrow \oneth$.

Apart from entering the relativistic fluid equations, the viscosities play another role for first order phase transitions.  They are prefactors in the nucleation rate \cite{nucleate}
\be
I = \frac{4}{\pi} \left( \frac{\sigma}{3T} \right)^{3/2}
\frac{\sigma (\zeta_{\rm H}+4\eta_{\rm H}/3) R_*}{\xi_{\rm H}^4 (\Delta w)^2}
e^{-\Delta F/T}
\ee

where $\Delta F = 4\pi\sigma R_*^2/3$ is the free energy and $R_*$ is the radius of a critical size fluctuation as given by the Laplace formula.  Also $\sigma$ is the surface free energy (surface tension) and $\Delta w$ is the difference in enthalpy densities of the two phases.  This is the probability per unit volume per unit time to nucleate an L (low temperature) phase bubble out of the H (high temperature) phase.  If one considers nucleating an H phase droplet in the L phase instead, one just needs to evaluate the correlation length and the viscosities in the L phase rather than the H phase.  At the critical temperature,
$R_* \rightarrow \infty$, and the rate vanishes because of the exponential.
Any system must supercool at least a minute amount in order that the rate attains a finite value.

It is important not only to ask what has RHIC told us about the equation of state, or what LHC will, but to make connections with other areas of physics such as cosmology and condensed matter physics.  To this end section 2 will summarize some interesting and important facts about atomic and molecular systems.  Relativistic dissipative fluid dynamics will be reviewed in section 3.  The shear viscosity will be studied in section 4 and the bulk viscosity in section 5.  The relevance and advances in applying gauge/gravity duality methods to transport coefficients in QCD are discussed in section 6.  An overview of possible experimental observables is given in section 7.  Concluding remarks are in section 8.  

\section{Atomic and Molecular Systems}

Shear viscosity $\eta$ is relevant for a change in shape at constant volume.  Bulk viscosity $\zeta$ is relevant for a change in volume at constant shape.  Here we review what is known about these in the vicinity of a phase transition in atomic and molecular systems.

In figures 1 through 3 we \cite{us} plot the ratio $\eta/s$ versus temperature at three fixed pressures, one of them being the critical pressure (meaning that the curve passes through the critical point) and the other ones being larger and smaller, for helium, nitrogen and water.  The ratio was constructed with data obtained from the National Institute of Standards and Technology (NIST) \cite{NIST}.  (Care must be taken to absolutely normalize the entropy to zero at zero temperature; we did that using data from CODATA \cite{CODATA}.)  The important observation \cite{singular} is that $\eta/s$ has a minimum at the 
critical point where there is a cusp.  At pressures below the critical pressure 
there is a discontinuity in $\eta/s$, and at pressures above it there is a broad 
smooth minimum. It is noteworthy that this behavior is exhibited by systems with very different phase transitions.  In the vicinity of the critical point, helium is a quantum system whereas nitrogen and water are classical systems.  Furthermore, nitrogen is a diatomic molecule whereas water is a polar molecule.  It is interesting to ask whether there is a universal scaling behavior. A plot of $(\eta/s)/(\eta/s)_c$ versus $T/T_c$, shown in figure 4, indicates that there is not.

The simplest way to understand the general behavior was 
presented by Enskog, as explained in \cite{atomtheory}.  Shear viscosity 
represents the ability to transport momentum.  In the classical transport theory of gases 
\be
\frac{\eta}{s} \sim T l_{\rm free} \bar{v} \sim \frac{T\bar{v}}{ n\sigma}
\ee
where $l_{\rm free}$ is the mean free path and $\bar{v}$ is the mean speed.  For a dilute gas the mean free path is large, $l_{\rm free} \sim 1/n\sigma$, with $n$ the particle number density and $\sigma$ the cross section.  Hence it is easy for a particle to carry momentum over great distances, leading to a large shear viscosity.  (This is the usual paradox, that a nearly ideal classical gas has a divergent shear viscosity.)  In a liquid there are strong correlations between neighboring atoms or molecules.  A liquid is homogeneous on a mesoscopic scale, but on a microscopic scale it is a mixture of clusters and voids.  The action of pushing on one atom is translated to the next one and so on until a whole row of atoms moves to fill a void, thereby transporting momentum over a relatively large distance and producing a large shear viscosity.  Reducing the temperature at fixed pressure reduces the density of voids, thereby increasing the shear viscosity.  The shear viscosity, normalized to the entropy, is observed to be the smallest at or near the critical temperature, corresponding to the most difficult condition to transport momentum.  This is an empirical observation.  It is also supported by numerical calculations.  For example, the shear viscosity was calculated for 2-dimensional Yukawa systems by Liu and Goree \cite{Goree}.  These have interesting applications to dusty plasmas and many other 2-dimensional condensed matter systems.  Figure 5 is a plot of the shear viscosity versus the Coulomb coupling parameter $\Gamma$.  Here $\Gamma = Q^2/aT$, where $Q$ is the charge and $a = 1/\sqrt{\pi n}$ is the Wigner-Seitz radius.  The shear viscosity is large at high T, which is a gas phase, where it is dominated by kinetic energy contributions.  It is also large at small T, which is a liquid phase, where it is dominated by potential energy contributions.  The shear viscosity has a minimum near $\Gamma = 17$ where kinetic and potential contributions contribute approximately equal amounts.

It has long been known that the bulk viscosity for a simple nonrelativistic gas of point particles, interacting via short-range 2-body forces, is vanishingly small.  This result has been attributed to Maxwell \cite{Maxwell}.  However, the bulk viscosity is not always small compared to the shear viscosity.  One quite relevant example is a diatomic gas.  The exchange of energy between translational and rotational degrees of freedom during collisions results in an appreciable bulk viscosity, which is important in the damping of sound waves \cite{atomtheory}.  

Another relevant example is a liquid-gas phase transition.  The bulk viscosity can be calculated from classical molecular dynamic simulations using the classical limit of the Kubo formula, which reads
\be
\zeta = \frac{1}{3T} \sum_{j=1}^3 \int_{-\infty}^0 dt \int d^3x
\langle \left( T_{ii}(0) - \langle T_{ii} \rangle \right)
\left( T_{jj}({\bf x},t) - \langle T_{jj} \rangle \right) \rangle
\label{classicalKuboShear}
\ee
where $T_{ij}$ are the spatial components of the energy-momentum tensor, and the angular brackets indicate the ensemble average.  This is just a measurement of the pressure fluctuations.  The results of one such simulation by Meier, Laesecke and Kabelac \cite{JChem} are shown in figure 6. The bulk viscosity (not normalized to the entropy density) shows a peak in the vicinity of, though not exactly at, the critical point.  The pressure fluctuations are large in this situation because there are two phases below the critical temperature with different energy densities.  The molecular dynamic simulations conserve energy, so this results in large fluctuations in the pressure.  An alternative point of view is to realize that two phases are in equilibrium at the same temperature and pressure but with different energy densities.  Hence at fixed $P$ and $T$ there can be large fluctuations in energy density.  As in the case of the diatomic gas, whenever it is easy to transfer energy between collective and internal degrees of freedom, one should expect an enhancement of the bulk viscosity.  The delay in attaining equilibrium due to energy exchange is ultimately responsible for entropy production.

Generally speaking, one ought to expect a minimum in the shear viscosity and a maximum in the bulk viscosity near a phase transition.  This is not a rigorous result but seems to be supported by various observations and calculations.  The dimensionless ratio of shear or bulk viscosity to entropy (disorder) would seem to be a good way to characterize the intrinsic ability of a substance to 
relax towards equilibrium independent of the actual physical conditions 
(gradients of pressure, energy density, etc.).  It would also seem to be a good way to compare very different substances.  Unfortunately, the entropy density is not always measured or calculated, or is not made available to the reader, at the same time as the viscosities.  This would facilitate more comparisons between atomic and molecular systems and strongly interacting matter at high temperature.

\section{Relativistic Dissipative Fluid Dynamics}

The energy-momentum tensor density for a perfect fluid (which does not imply 
that the matter is non-interacting) is $T^{\mu\nu} = -
Pg^{\mu\nu}+wu^{\mu}u^{\nu}$.  Here $w=P+\epsilon=Ts$ is the local enthalpy 
density and $u^{\mu}$ is the local flow velocity.  Corrections to this 
expression are proportional to first derivatives of the local quantities whose 
coefficients are the shear viscosity $\eta$, bulk viscosity $\zeta$, and thermal conductivity $\chi$.  Explicit expressions may be found in textbooks \cite{fluid1,fluid2} which are useful to summarize here.  Dissipative contributions are added to the energy-momentum tensor and baryon current as follows.
\ba
T^{\mu\nu} &=& -P g^{\mu\nu} + w u^{\mu}u^{\nu} + \Delta T^{\mu\nu}
\nonumber \\
J_{\rm B}^{\mu} &=& n_{\rm B} u^{\mu} + \Delta J_{\rm B}^{\mu}
\ea
There are two common definitions of the flow velocity in relativistic dissipative fluid dynamics which are important to distinguish.

In the Eckart approach $u^{\mu}$ is the velocity of baryon number flow.  The dissipative terms must satisfy the conditions
$\Delta J^{\mu}_{\rm B}=0$ and 
$u_{\mu} u_{\nu} \Delta T^{\mu\nu} = 0$, the latter following from the 
requirement that $T^{00}$ be the energy density in the local (baryon) rest 
frame.  The most general form of $\Delta T^{\mu\nu}$ is
\be
\Delta T^{\mu\nu} = 
\eta \left(\Delta^{\mu} u^{\nu} + \Delta^{\nu} u^{\mu}\right)
+\left(\twoth \eta - \zeta\right) H^{\mu\nu} \partial_{\rho} u^{\rho}
- \chi \left( H^{\mu\alpha} u^{\nu} + H^{\nu\alpha} u^{\mu}
\right) Q_{\alpha} \, .
\ee
Here
\be
H^{\mu\nu} = u^{\mu} u^{\nu} - g^{\mu\nu}
\ee
is a projection tensor normal to $u^{\mu}$,
\be
\Delta_{\mu} = \partial_{\mu} - u_{\mu} u^{\beta} \partial_{\beta}
\ee
is a derivative normal to $u^{\mu}$, and
\be
Q_{\alpha} = \partial_{\alpha} T - T u^{\rho} \partial_{\rho} u_{\alpha}
\ee
is the heat flow vector whose nonrelativistic limit is ${\bf Q} = 
- \mbox{\boldmath $\nabla$} T$.  The entropy current is
\be
s^{\mu} = s u^{\mu} + \frac{1}{T} u_{\nu} \Delta T^{\mu\nu} \, .
\ee
Its divergence is
\ba
\partial_{\mu}s^{\mu} &=& \frac{\eta}{2T}
\left( \partial_iu^j + \partial_ju^i - \twoth \delta^{ij} \nabla \cdot {\bf u}
\right)^2 \nonumber \\
& & \mbox{} + \frac{\zeta}{T} \left( \nabla \cdot {\bf u} \right)^2
+ \frac{\chi}{T^2} \left( \nabla T + T \dot{{\bf u}}\right)^2 \, .
\ea
All three dissipation coefficients must be non-negative to insure that entropy 
can never decrease.

In the Landau-Lifshitz approach $u^{\mu}$ is the velocity of energy transport.  
The dissipative part of the energy-momentum tensor satisfies $u_{\mu} \Delta 
T^{\mu\nu} = 0$, and $\Delta J_{\rm B}^{\mu}$ is not constrained to be zero.
In this case the most general form of the energy-momentum tensor is
\be
\Delta T^{\mu\nu} = 
\eta \left(\Delta^{\mu} u^{\nu} + \Delta^{\nu} u^{\mu}\right)
+\left(\twoth \eta - \zeta\right) H^{\mu\nu} \partial_{\rho} u^{\rho} \, .
\ee
The baryon current is modified to
\be
\Delta J^{\mu}_{\rm B} = \chi \left(\frac{n_{\rm B}T}{w}\right)^2
\Delta^{\mu} \left(\frac{\mu_B}{T}\right) \, .
\ee
Even though the entropy current in this approach is different, being
\be
s^{\mu} = s u^{\mu} - \frac{\mu_{\rm B}}{T} \Delta J_{\rm B}^{\mu} \, ,
\ee
its divergence is the same.  Physical, observable results cannot depend on how one defines the frames of reference.

In nuclear collisions at RHIC the ratio of the net baryon number to the total number of particles is very small, typically less than 1\%.  With a greater energy available at the LHC it will be smaller yet.  Typically hydrodynamic calculations set the net baryon number to zero as a good approximation.  In that case one has no choice but to use the Landau-Lifshitz definition of flow velocity.  Then thermal conductivity is neither relevant nor defined.

First order dissipative fluid dynamics applies when the viscosities are small, or when the gradients are small, or both.  The dispersion relations for the transverse and longitudinal (pressure) parts of the momentum density are
\ba
\omega + i D_t k^2 &=& 0 \nonumber \\
\omega^2 - v_s^2 k^2 + i D_l \omega k^2 &=& 0
\ea
where $D_t = \eta/w$ and $D_l = (\fourth \eta + \zeta)/w$ are diffusion 
constants with the dimension of length and $v_s$ is the speed of sound.  Since 
$w=Ts$, the dimensionless ratio of shear or bulk viscosity to entropy density 
is a good way to characterize the strongly interacting matter created at RHIC or LHC and to compare it to ordinary atomic and molecular systems. 

\section{Shear Viscosity}

What are the theoretical predictions for the shear viscosity of hadrons and quark-gluon plasma?  In the low energy chiral limit for pions the cross section is proportional to $\hat{s}/f_{\pi}^4$, where $\hat{s}$ is the usual Mandelstam variable for invariant mass-squared and $f_{\pi}$ is the pion decay constant.  The thermally averaged cross section is $\langle \sigma \rangle \propto T^2/f_{\pi}^4$, which leads to $\eta/s \propto (f_{\pi}/T)^4$.  Explicit calculation gives \cite{Prakash}
\be
\frac{\eta}{s} \approx \frac{15}{16\pi} \frac{f_{\pi}^4}{T^4}
\label{chiral}
\ee
Thus the ratio $\eta/s$ diverges as $T \rightarrow 0$.  At the other extreme 
lies quark-gluon plasma.  The parton cross section behaves as $\sigma \propto 
g^4/\hat{s}$.  A first estimate yields $\eta/s \propto 1/g^4$.  Asymptotic 
freedom at one loop order gives $g^2 \propto 1/\ln(T/\Lambda_T)$ where 

$\Lambda_T$ is proportional to the scale parameter $\Lambda_{\rm QCD}$ of QCD.  
Therefore $\eta/s$ is an increasing function of $T$ in the quark-gluon phase.  
As a consequence, $\eta/s$ must have a minimum.  Based on atomic and molecular 
data, this minimum should lie at the critical temperature if there is one, 
otherwise at or near the rapid crossover temperature.

The most accurate and detailed calculation of the viscosity in the low 
temperature hadron phase was performed in \cite{Prakash}.  (See also \cite{Davesne}.)  The two-body interactions used went beyond the chiral approximation, and included intermediate resonances such as the $\rho$-meson.  To calculate the entropy density a free gas consisting of pions, $\eta$ mesons, kaons, $\rho$ mesons, $\omega$ mesons, and $K^*$ mesons was used.  The results are displayed in figure 7, both two flavors (no kaons or $K^*$) and three flavors (with kaons and $K^*$).  The qualitative behavior is the same as in eq. (\ref{chiral}).  

The most accurate and detailed calculation of the shear viscosity in the high temperature quark-gluon phase was performed in \cite{Arnold}.  They used perturbative QCD to calculate the full leading-order expression, including summation of the Coulomb logarithms.  For three flavors of massless quarks the result is
\be
\frac{\eta}{s} = \frac{5.12}{g^4 \ln(2.42/g)}
\ee
Use this together with the two-loop renormalization group expression for the 
running coupling
\be
\frac{1}{g^2(T)} = \frac{9}{8\pi^2} \ln\left( \frac{T}{\Lambda_T} \right)
+ \frac{4}{9\pi^2} \ln \left( 2 \ln\left( \frac{T}{\Lambda_T} \right) \right)
\ee
with $\Lambda_T = 30$ MeV, which approximately corresponds to using an energy 
scale of $2\pi T$ and $\Lambda_{\overline{MS}} = 200$ MeV.  The result is also 
plotted in figure 7 \cite{us}.  These results imply a minimum in the neighborhood of the expected value of $T_c \approx 190$ MeV.  Whether there is a discontinuity or a smooth crossover cannot be decided since both calculations are unreliable near $T_c$.

It is interesting to ask what happens in the large $N_c$ limit with $g^2N_c$ 
held fixed \cite{largeN}.  In this limit, meson masses do not change very much 
but baryon masses scale proportional to $N_c$; therefore, baryons may be 
neglected in comparison to mesons due to the Boltzmann factor.  Since the meson 
spectrum is essentially unchanged with increasing $N_c$, so is the Hagedorn 
temperature.  The critical temperature to go from hadrons to quarks and gluons 
is very close to the Hagedorn temperature, so that $T_c$ is not expected to 
change very much either.  In the large $N_c$ limit the meson-meson cross section 
scales as $1/N_c^2$.  According to our earlier discussion on the classical 
theory of gases, this implies that the ratio $\eta/s$ in the hadronic phase 
scales as $N_c^2$.  This general result is obeyed by (\ref{chiral}) since it is 
known that $f_{\pi}^2$ scales as $N_c$.  The large $N_c$ limit of the viscosity 
in the quark and gluon phase may be inferred from the calculations of 
\cite{Arnold} to be
\be
\left(\frac{\eta}{s}\right)_{\rm QGP} = 
\left( \frac{1+3.974r}{1+1.75r} \right)
\frac{69.2}{\left(g^2N_c\right)^2 \ln \left( 26/(g^2N_c(1+0.5r))\right)}
\ee
where $r = N_f/N_c$.  Thus the ratio $\eta/s$ has a finite large $N_c$ limit in 
the quark and gluon phase.  Therefore, we conclude that $\eta/s$ has a 
discontinuity proportional to $N_c^2$ if $N_c \rightarrow \infty$.  This jump is 
in the opposite direction to that in figure 7.

There are other means to compute the shear viscosity.  Why not just compute it from numerical calculations with lattice gauge theory?  The basic problem, which has been known for 25 years, is that one must determine the spectral density in frequency space from a Fourier transform of a function in (imaginary) time whose values are only known at a finite number of steps, typically $N_{\tau} = 4, 8, 16, 32$, and then find the zero frequency limit.  For example, quantitative results for the shear viscosity in lattice gauge theory have been reported by Nakamura and Sakai \cite{latticeeta} for pure SU(3) without quarks.  (This theory has a first order phase transition.)  This bold effort obtained $\eta/s \approx 0.5$ in the temperature range $1.6 < T/T_c < 2.2$, albeit with uncertainties of order 100\%.  An improved calculation by Meyer \cite{Meyershear,Meyerreview} obtained $\eta/s \approx 0.134$ at $T/T_c = 1.65$ and $\eta/s \approx 0.102$ at $T/T_c = 1.24$.  Taken together these suggest a minimum at or near $T_c$.

Another interesting approach has been the computation of $\eta/s$ in 
${\cal N}=4$ supersymmetric $SU(N_c)$ Yang-Mills theory (SYM).  In that theory it is possible to do calculations in the large coupling limit \cite{SUSY} and in the weak coupling limit \cite{Huot}, and without too much imagination it is possible to find a paramtrization that interpolates smoothly between the two limits.  However, SYM has no renormalization group running coupling, no asymptotic freedom, and no thermodynamic phase transition.  Since it has so many more degrees of freedom than QCD as possible scattering targets, its viscosity to entropy ratio is much smaller than QCD at high temperature when compared at the same value of the gauge coupling.  However, when the theories are compared at the same value of the Debye screening mass they do agree reasonably well \cite{Huot}.

In another approach, Gelman, Shuryak and Zahed \cite{cQCD} have modeled the dynamics of long wavelength modes of QCD at temperatures from $T_c$ to $1.5 T_c$ as a classical, nonrelativistic gas of massive quasi-particles with color charges.  They obtained a ratio of $\eta/s \approx 0.34$ in this temperature range.  One can also perform hadronic molecular dynamic simulations of hadronic collisions within a box to compute the shear viscosity from eq. (\ref{classicalKuboShear}).  This has been done by Muronga \cite{MurongaTransport} using the well-known UrQMD (Ultra-relativistic Quantum Molecular Dynamics) program which has been used so successfully for heavy ion collisions at energies below RHIC and as an afterburner in RHIC collisions.  The practical problem with this approach is that reactions of the type $2 \rightarrow n$ particles are fairly well-known, and can be implemented in the computer simulation, but how to handle the inverse process $n \rightarrow 2$ particles is not.  Therefore all reactions involving $n > 2$ had to be turned off, otherwise detailed balance could not be achieved.  The results were within a factor of 2 to 4 compared to those of Prakash {\it et al.} \cite{Prakash}, depending on temperature.

\section{Bulk Viscosity}

What are the theoretical predictions for the bulk viscosity of hadrons and quark-gluon plasma?  In the low energy chiral limit for pions the equation of state is $\epsilon = 3 P$ and $v_s^2 = 1/3$.  Hence the bulk viscosity vanishes in the low temperature limit.  However, there are deviations from this conformal behavior due to interactions among the pions which involve a characteristic scale $\Lambda_p \approx 275$ MeV.  Explicit calculation gives \cite{ChenWang}
\be
\frac{\zeta}{s} \approx \frac{9}{8\pi^2} \left( \ln\frac{\Lambda_p}{T} - \frac{1}{4} \right) \left( \ln\frac{\Lambda_p}{T} - \frac{3}{8} \right)
\frac{T^4}{f_{\pi}^4}
\label{chiralbulk}
\ee
The ratio $\zeta/s$ goes to zero as $T \rightarrow 0$, in contrast to $\eta/s$ which diverges.  

At the other extreme lies quark-gluon plasma.  At asymptotically high temperatures, much greater than quark masses and the QCD scale parameter $\Lambda_{\overline{MS}}$, the system is again essentially conformally invariant and the bulk viscosity should go to zero. The most accurate and detailed calculation of the bulk viscosity in the high temperature 
quark-gluon phase was performed in \cite{Arnoldbulk}.  For three flavors of massless quarks the result is
\be
\frac{\zeta}{s} = \frac{g^4}{5000 \ln(6.34/g)} \, .
\ee
There are two significant differences compared to the corresponding expression for the shear viscosity.  First, the overall coefficient is four orders of magnitude smaller, and second, the factor $g^4$ appears in the numerator rather than in the denominator.  The ratio $\zeta/s$ goes to zero at high temperature because the strong coupling $g^2(T)$ goes to zero.  Conversely it becomes large at low temperatures whose scale is set by $\Lambda_{\overline{MS}}$.

The picture painted by massive versus massless pions is somewhat different.  Figure 9 shows the results from \cite{Prakash} (see also \cite{Davesne}), which calculated the bulk viscosity from the elastic scattering of massive pions using experimentally inferred cross sections.  As for the shear viscosity, the entropy density was calculated on the basis of a free gas of pions and $\eta$, $\rho$ and $\omega$ mesons.  Contrary to massless pions, the ratio $\zeta/s$ is a decreasing function of $T$ rather than an increasing function of $T$.  From figure 9 one might conclude that $\zeta/s$ is a monotonically decreasing function of $T$ even as one goes from the hadronic phase to the plasma phase.  However, the figure is also consistent with a peak in the vicinity of 150 to 200 MeV.

Indeed, other approaches strongly suggest a peak in $\zeta/s$ in the vicinity of the phase transition or rapid crossover.  Using a combination of an exact sum rule derived from the low-energy theorems of broken scale invariance in QCD, lattice data on the equation of state which sees a large increase in 
$\epsilon-3P$ in the vicinity of $T_c$, and an ansatz for the correlation function of the energy-momentum tensor, Kharzeev and collaborators have estimated the behavior of $\zeta/s$ above $T_c$ \cite{Kharzeev1,Kharzeev2}.  Their result is that $\zeta/s$ goes to zero rapidly at high temperature but becomes quite large as $T_c$ is approached from above.  For example, 
$\zeta/s < 0.01$ for $T/T_c > 1.4$, 
$\zeta/s \approx 0.025$ at $T/T_c = 1.2$, $\zeta/s \approx 0.07$ at $T/T_c = 1.1$, and $\zeta/s \approx 0.3$ at $T/T_c = 1.01$.  The large $T$ limit is numerically consistent with the perturbative QCD results displayed in figures 8 and 9, and shows a very strong growth as $T_c$ is approached.  However, it should be mentioned that Moore and Saremi have challenged this analysis \cite{Saremi}.

A direct calculation of $\zeta/s$ has been done by Meyer \cite{Meyerbulk,Meyerreview} using lattice methods.  Such a direct approach avoids the necessity to make an ansatz for the spectral density, but on the other hand it is difficult to access the low frequency limit of the spectral density.  The results for pure gauge theory without quarks (this theory has a first order phase transition) are as follows: $\zeta/s < 0.015$ at $T/T_c = 3.2$, $\zeta/s \approx 0.008$ at $T/T_c = 1.65$, and $\zeta/s \approx 0.065$ at $T/T_c = 1.24$.  The latter two points were obtained with a temporal lattice size of $N_{\tau} = 12$.  For a smaller lattice of $N_{\tau} = 8$, the result was $\zeta/s \approx 0.73$ at $T/T_c = 1.02$.  Taking into account systematic and statistical uncertainties, both of these approaches agree with each other, from which we conclude that a sharp rise is predicted by lattice gauge theory as $T_c$ is approached from above.

Just as the ability to exchange energy between translational and internal degrees of freedom increases the bulk viscosity, so does the ability to store energy in, and extract energy from, a mean field.  A simple formula was derived by Paech and Pratt \cite{Paech} for the ubiquitous example of a $\sigma$ field.
\be
\frac{\zeta}{s} = \frac{\Gamma}{m_{\sigma}^2} 
\left( \frac{\partial P}{\partial \sigma} \right)_{\epsilon}
\frac{\partial \sigma_{\rm eq}}{\partial s}
\ee
Here the derivative of the pressure with respect to the mean field is carried out at fixed energy density.  The damping coefficient is obtained from the dispersion relation of the field at zero momentum.
\be
\omega^2 = m_{\sigma}^2 - i \Gamma \omega
\ee
Whenever the field $\sigma$ is rapidly varying or when $m_{\sigma}$ is small or both, such as near a phase transition, the bulk viscosity is predicted to be large.  Within a simple mean-field model they found the behavior near the critical temperature of a first-order phase transition to be
\be
\frac{\zeta}{s} \sim |T-T_c|^{-n}
\ee
with $n=1$.  In a more sophisticated approach the critical exponent is expected to be somewhat different.  Whether the ratio $\zeta/s$ diverges or not in full QCD is an interesting and open question \cite{Kharzeev2}.

\section{Gauge/Gravity Correspondence}

The AdS/CFT (Anti-de Sitter/Conformal Field Theory) correspondence, or more generally gauge/gravity duality, may offer new and profound insights into the non-perturbative realm of QCD \cite{Mald,Witten,Polyakov}.  The bottom-up approach, referred to as AdS/QCD, constructs a dual gravity theory with just a few parameters which are fit to some set of hadronic data.  Then all other observables can be predicted.  This approach has been of particular interest for calculating transport coefficients ever since the influential paper of Kovtun, Son and Starinets \cite{Son}.  The basic ideas and results are summarized here; for a more extensive introduction to this subject, see \cite{SS}.  

Consider a background metric in 5-dimensional space of the generic form
\be
ds^2 = g_{00}(r) dt^2 + g_{xx}(r) |d{\bf x}|^2 + g_{rr}(r) dr^2
\label{backgroundg}
\ee
where the three functions $g_{00}$, $g_{xx}$ and $g_{rr}$ depend on the fifth dimensional variable $r$ only.  The convention is that $g_{00} = - 1$ and $g_{xx} = 1$ in flat 4-dimensional space.  Our normal 4-dimensional world exists at $r \rightarrow \infty$.  There is a black brane (in four dimensions it would be a black hole) horizon located at $r = r_0$.  As the horizon is approached it is assumed that the metric behaves as follows: $g_{rr} \rightarrow \gamma_r/(r-r_0)$ and $g_{00} \rightarrow - \gamma_0 (r-r_0)$ while $g_{xx}$ approaches the finite value $g_{xx}(r_0)$.  The Hawking temperature is
\be
T = \frac{1}{4\pi} \sqrt{\frac{\gamma_0}{\gamma_r}} \, .
\ee
It is this black brane that provides for a finite temperature system.
The matter and radiation fields which generate this metric are oftentimes not known.  The procedure then is to compute the correlation functions of the energy-momentum tensor to infer the long wavelength dispersion relations from (i) the poles of these functions or (ii) the Kubo formulas, (iii) to use the membrane paradigm where one solves the field equations with boundary conditions applied on a stretched horizon of the black brane, or (iv) to solve for gauge invariant perturbations with approriate conditions.  The results from all of these methods coincide, as demonstrated for cases (i), (ii) and (iv) in \cite{KS}.  See \cite{LiuIqbal} for case (iii). 

A quite general formula for the ratio of shear viscosity to entropy density was first derived in \cite{stretched} and reproduced by different methods in \cite{Springer1}.
\be
\frac{\eta}{s} = T \, \frac{\sqrt{-g(r_0)}}{\sqrt{-g_{00}(r_0) g_{rr}(r_0)}}
\int_{r_0}^{\infty} dr \frac{-g_{00}(r) g_{rr}(r)}{\sqrt{-g(r)} g_{xx}(r)} \, .
\label{genshear}
\ee
Hence, knowing the background metric one may compute $\eta/s$ just by evaluating this integral.  A frequently used metric is one that is dual to ${\cal N} = 4$ SU($N$) supersymmetric Yang Mills theory at finite temperature and in the limit $N \rightarrow \infty$ and $g^2 N \rightarrow \infty$,
\be
ds^2 = \frac{r^2}{L^2} \left[ -f(r)dt^2 + |d{\bf x}|^2 \right]
 + \frac{L^2}{r^2 f(r)} dr^2 \, .
\label{SUSY}
\ee   
Here $f(r) = 1-(r_0/r)^4$.  Using this metric yields $\eta/s = 1/4\pi$, which is the famous result.  In fact, there is a theorem which states that whenever
$R^0_{\, 0} = R^x_{\, x}$, where $R_{\mu\nu}$ is the Ricci tensor, then $\eta/s = 1/4\pi$ \cite{Buchel1}.  This equality of components of the Ricci tensor is true whenever the background metric is generated by scalar fields \cite{Springer2}.
  
The gravity dual to QCD is not known; in fact, it is not even known if one exists.  AdS/CFT can only be an approximate representation of QCD at very high energies or temperatures.  The reason is that a conformal field theory has no intrinsic energy scale, although it may have dimensionless parameters.  This is approximately representative of QCD at high energies or temperatures since the QCD gauge coupling varies only logarithmically.  There are several phenomenological models for QCD that incorporate confinement.  The simplest of these is the hard wall model \cite{hardwall} which uses the metric of Eq. (\ref{SUSY}) but with $r$-space cutoff at some value $r_{\rm min}$.  This leads to radial excitations of the vector and axial-vector meson spectra with the mass being linear in the radial quantum number $n$.  Introduction of the new scale $r_{\rm min}$ implies that the transport coefficients might vary nontrivially with temperature $T$.

The soft wall model was developed to improve upon the hard wall model.  In particular, it leads to linear Regge trajectories wherein the radial excitations of the vector and axial-vector meson spectra have mass-squared being linear in the radial quantum number $n$, in substantial agreement with data.  
It can be obtained by adding a dilaton field to the usual AdS$_5$ metric to break the scale invariance.  The dilaton profile which leads to Regge behavior is $\phi(r) = c L^4/r^2$, where $c$ is a constant which can be determined by fitting the meson spectrum.  If one computes the ratio $\eta/s$ using the soft wall model of \cite{softwall} and Eq. (\ref{genshear}), the result is $1/4\pi$ because the metric is exactly AdS$_5$.  There is an alternative formulation of the soft wall.  Instead of adding a nontrivial dilaton, one keeps the dilaton constant while deforming the metric away from AdS$_5$.  This version of the soft wall model has been studied in \cite{Andreev}.  The deformed metric is
\be 
d \hat{s}^2 = e^{-2\phi(r)}\left[\frac{r^2}{L^2} \left( -f(r)dt^2 + |d{\bf x}|^2 \right) + \frac{L^2}{r^2 f(r)} dr^2 \right] \, .
\label{softwallmetric}
\ee
It should be stressed that this is a different implementation of the soft wall model than that of \cite{softwall}.  Both pictures lead to linear Regge trajectories, but other physical quantities may differ.  Using this metric in Eq. (\ref{genshear}) yields
\be
\frac{\eta}{s} = \frac{1}{2\pi x_0}\left[ 1+\frac{1}{x_0}
\left(e^{-x_0} -1 \right) \right] 
\ee
where
\be
x_0 = \frac{3 c}{\pi^2 T^2} \, .
\ee
The Hawking-Page analysis of the phase transtition in \cite{softwallD} gives $T_c = 1/(2\sqrt{c}) = m_{\rho}/4 \approx 192$ MeV.  Note that the conformal limit of $1/4\pi$ is approached from below as $T$ increases.  One may be surprised that the shear viscosity is not greater than or equal to $1/4\pi$.  The reason is that the theorem of \cite{Buchel1} is not applicable because $R^0_{\, 0} \neq R^x_{\, x}$.  Instead, in this version of the soft wall model, one finds by explicit computation that
\be
R^0_{\, 0} - R^x_{\, x} = -6 \phi^{\prime}(r)e^{2\phi(r)}\frac{r_0^4}{r^3 L^2}
\ee  
which is nonzero as long as $\phi(r)$ is not constant.  This may be a sign that a metric such as (\ref{softwallmetric}) cannot be generated by conventional supergravity matter fields.  It remains an open question.

As discussed earlier, the bulk viscosity vanishes for conformally invariant theories, in particular AdS/CFT.  So far, there is no known formula for the bulk viscosity analogous to Eq. (\ref{genshear}).  It seems necessary to know what fields generate the background metric.  The simplest case is when the metric is created by a single scalar field $\phi$ with action
\be
S = \frac{1}{16\pi G_5} \int d^5x \sqrt{-g} \left[ R
- \thalf \partial^{\mu} \phi \partial_{\mu} \phi - V(\phi) \right]
\ee
where $G_5$ is the gravitational constant in five dimensions, $R$ is the curvature, and $V(\phi)$ is a potential.  Examples of such potentials are given by Chamblin and Reall \cite{Chamblin}
\be
V(\phi) = V_0 {\rm e}^{\gamma \phi}
\ee
and by Gubser {\it et al.} \cite{Gubsers}
\be
V(\phi) = - \frac{12}{L^2} \cosh \left(\gamma \phi \right)
+ b \phi^2 
\ee
where $V_0$, $\gamma$, and $b$ are constants.  These and related potentials have parameters that can be varied to approximate the QCD equation of state as obtained from lattice QCD calculations.  For example, for the Chamblin-Reall potential the speed of sound is
\be
v_s^2 = \frac{1}{3} - \frac{\gamma^2}{2} \, .
\ee
In the absence of a top-down approach it is very difficult to know which potential provides the best representation of QCD.

Buchel \cite{Buchel2} has conjectured that there is a minimum value of the bulk viscosity relative to the shear viscosity on the basis of a variety of holographically dual computations.  This inequality is
\be
\frac{\zeta}{\eta} \ge 2 \left( \oneth - v_s^2 \right) \, .
\ee
Rather than become engrossed in detailed models, let us take a more general point of view to see what can be learned.

Motivated by the above studies, Springer \cite{Springer2} considered the following constraint on the metric
\be
g_{00}(r) = \frac{a_0}{a_2-3}g_{xx}(r)+a_1 g_{xx}^{a_2-2}(r) \, .
\ee
The requirement that $ R^0_{\, 0} = R^x_{\, x}$ leads to
\be
g_{rr}(r) = \frac{-g_{00}(r) g^3_{xx}(r)}{a^2_3}
\left( \frac{d}{dr} \ln \left[g_{00}(r) g^{xx}(r) \right] \right)^2 \, .
\ee
The parameter $a_1$ is not independent but is fixed in terms of the others by the requirement that $g_{rr} \rightarrow \gamma_r/(r-r_0)$ and $g_{00} \rightarrow - \gamma_0 (r-r_0)$ while $g_{xx}$ approaches the finite value $g_{xx}(r_0)$.  The Hawking temperature is
\be
T = \frac{a_3}{4\pi} g^{-3/2}_{xx}(r_0) \, .
\ee
The speed of sound and the viscosities can be found by calculating the dispersion relation.
\ba
v_s^2 &=& \frac{a_0 -a_2}{3} \nonumber \\
\frac{\eta}{s} &=& \frac{1}{4\pi} \nonumber \\
\frac{\zeta}{s} &=& \frac{a_2 - 1}{3\pi}
\ea
These results respect the conjectured minimum value of $\zeta/\eta$ if $a_0 + a_2 \ge 3$; the equality is attained in certain models.  By specifying the metric one may deduce the speed of sound (hence the equation of state) and the bulk viscosity.  The advantage of this approach is that it allows one to correlate the equation of state and the dissipative coefficients in a consistent way.  The disadvantage is that until we have an approximate or exact gravity dual to QCD, this approach will remain phenomenological.

\section{Observable Consequences}

It ought to be possible to extract numerical values of the viscosities in heavy ion collisions via scaling violations to perfect fluid flow predictions \cite{Bonasera}.  The program is to solve relativistic viscous fluid equations, with appropriate initial conditions and with a hadron cascade afterburner \cite{burner}, over a range of beam energies and nuclei and extract $\eta(T)/s(T)$ and $\zeta(T)/s(T)$ from comparison with data.  This program is analogous to what was accomplished at lower energies of 30 to 1000 MeV per nucleon beam energies in the lab frame.  At those energies, it was possible to infer the compressibility of nuclear matter and the momentum-dependence of the nuclear optical potential via the transverse momentum distribution relative to the reaction plane \cite{BUU} and via the balance between attractive and repulsive scattering \cite{balance}.

To make quantitative comparisons with data it is necessary to solve relativistic viscous fluid equations numerically in three space dimensions without assuming any particular symmetry.  On general grounds one should expect viscous effects to smooth out gradients in temperature and flow velocity and to slow down the expansion of the system.  To develop some intuition as to what happens it is worthwhile to solve those equations in one space and one time direction under the assumption of boost invariance \cite{McGill2007}.  This is the fluid dynamical model proposed by Bjorken \cite{Bjorken}.

For the sake of simplicity, consider a bag model type equation of state at temperatures above the critical one.  The pressure is  
\be
P = N_{\rm dof}\frac{\pi^2}{90}T^4 - B
\ee
and the energy and entropy densities follow directly.  Neglect the shear viscosity in order to focus attention on the bulk viscosity.  The fluid dynamic equation to solve is
\be
\frac{d\epsilon}{dt} + \frac{\epsilon + P}{t} - \frac{\zeta}{t^2} = 0
\label{BJviscous}
\ee
where $t$ denotes the local time (we do not use the symbol $\tau$ as is customarily done so as not to confuse it with the relaxation time discussed earlier).  First make the assumption that $\zeta$ is a monotonically decreasing function of $T$ to the inverse power of $n > 0$.
\be
\zeta(T) = \zeta_i \left( \frac{T_i}{T} \right)^n
\ee
Here the subscript $i$ indicates the initial value at the beginning of the fluid expansion.  The solution is easily found.
\be
\left( \frac{T(t)}{T_i} \right)^{n+4} = \left( \frac{t_i}{t} \right)^{(n+4)/3}
\left[ 1 - \frac{n+4}{n+1} \frac{\zeta_i}{s_i} \frac{1}{t_i T_i} \right]
+ \frac{n+4}{n+1} \frac{\zeta_i}{s_i} \frac{1}{t_i T_i} \left( \frac{t_i}{t} \right)
\ee
The second term in the square bracket should be small comapared to one in order that the gradients are not so large initially as to negate the application of viscous fluid dynamics.  The last term on the right hand side shows that the temperature decreases with time more slowly in the presence of the bulk viscosity.  In fact that term dominates at large $t$.  Eventually the temperature will fall to $T_c$ and one would need to switch to the hadronic equation of state and associated viscosity.

Next make the assumption that $\zeta$ has a power singularity at $T_c$.
\be
\zeta(T) = \zeta_i \left( \frac{T_i - T_c}{T-T_c} \right)^n
\ee
The equation of motion cannot be solved in terms of elementary functions.  However, the asymptotic behavior can be determined to be
\be
T(t) \rightarrow T_c + (T_i - T_c) \left( \frac{\zeta_i}{s_i} \right)^{1/n} 
\left( \frac{1}{t T_c} \right)^{1/n} \;\;\; {\rm as}\; t \rightarrow \infty \, .
\ee
It takes an infinite amount of time to reach the critical temperature.  This is an example of critical slowing down.  Of course this would not happen in the real world, because eq. (\ref{BJviscous}) neglects transverse expansion and the surface emission of particles near the cooler periphery of the matter.

At RHIC and LHC some of the specific proposals to extract or infer the viscosity to entropy ratio from data include: elliptic flow \cite{Teaney}, Hanbury-Brown and Twiss (HBT) interferometry \cite{Teaney}, single particle momentum spectra \cite{Teaney,Baier}, and momentum fluctuations \cite{Gavin}.  Other possibilities include jet quenching and photon and dilepton spectra.  Some of the complications include the possibility that gradients are so large that the second-order dissipative equations of Israel and Stewart are necessary \cite{Muronga} and that turbulence in the plasma may lead to an anomalous viscosity \cite{Asakawa}.  Clearly this is an interesting and challenging goal but worth the effort.

\section{Conclusion}

There are strong arguments and calculations which suggest that 
hadron/quark-gluon matter should have a {\it minimum} in the ratio of shear viscosity to entropy, and a {\it maximum} in the ratio of bulk viscosity to entropy, at or near the critical or crossover point in the phase diagram.  Suffiently detailed calculations and measurements ought to allow us to infer these quantities from experiments at RHIC and LHC.  These are interesting dimensionless measures of dissipation relative to disorder which can be compared among a wide variety of substances.  

My conclusion is that RHIC is a {\it thermometer} since it measures hadron ratios and photon and dilepton spectra, it is a {\it barometer} since it measure radial and elliptic flow, and it may be a {\it viscometer} since it could measure deviations from ideal (nonviscous) fluid flow.  Soon the same will happen with the LHC.  There is much work still to be done by both theorists and experimentalists!

\section*{Acknowledgements}

This work is based to some extent on \cite{us,QM2006,McGill2007}.  I wish to thank L. P. Csernai, L. D. McLerran, M. Prakash, P. Chakraborty, S. Gavin, T. Springer, and O. Andreev for past and present dicussions.  This work was supported by the US Department of Energy under grant DE-FG02-87ER40328.

\newpage

\begin{figure}
 \centering
 \includegraphics[width=3.5in,angle=90]{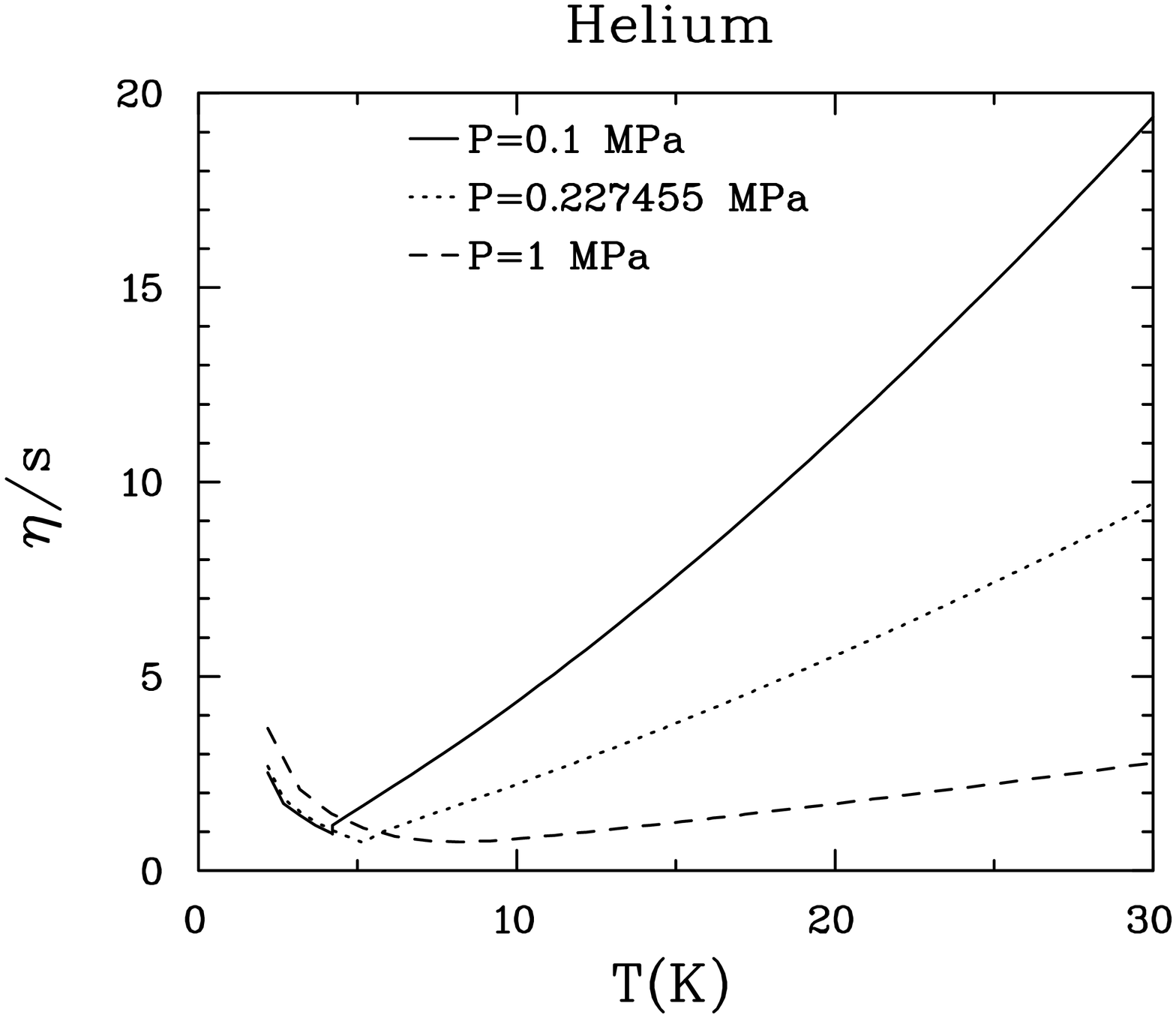}
 \caption{The ratio $\eta/s$ as a function of $T$ for helium with $s$ normalized 
such that $s(T=0)=0$.  The curves correspond to fixed pressures, one of them 
being the critical pressure, and the others being greater (1 MPa) and the other 
smaller (0.1 MPa).  Below the critical pressure there is a jump in the ratio, 
and above the critical pressure there is only a broad minimum. They were 
constructed using data from NIST and CODATA.  From Csernai, Kapusta and McLerran.}
 \label{fighelium}
\end{figure}

\begin{figure}
 \centering
 \includegraphics[width=3.5in,angle=90]{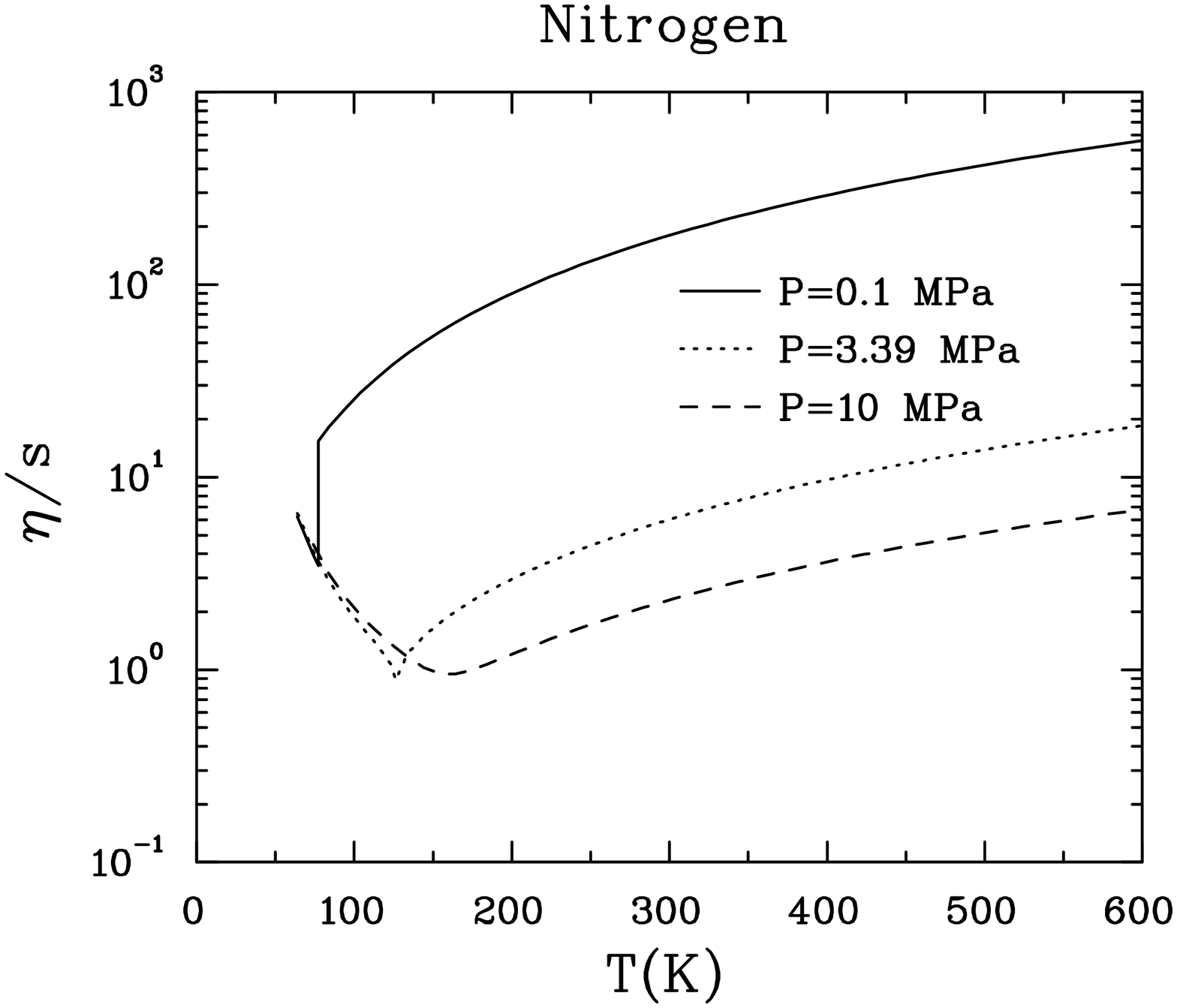}
 \caption{The ratio $\eta/s$ as a function of $T$ for nitrogen with $s$ 
normalized such that $s(T=0)=0$.  The curves correspond to fixed pressures, one 
of them being the critical pressure, and the others being greater (10 MPa) and 
the other smaller (0.1 MPa).  Below the critical pressure there is a jump in the 
ratio, and above the critical pressure there is only a broad minimum. They were 
constructed using data from NIST and CODATA.  The curves are plotted on 
logarithmic scale to make the behavior around the critical point more visible.  From Csernai, Kapusta and McLerran.}
 \label{fignitrogen}
\end{figure}

\begin{figure}
 \centering
 \includegraphics[width=3.5in,angle=90]{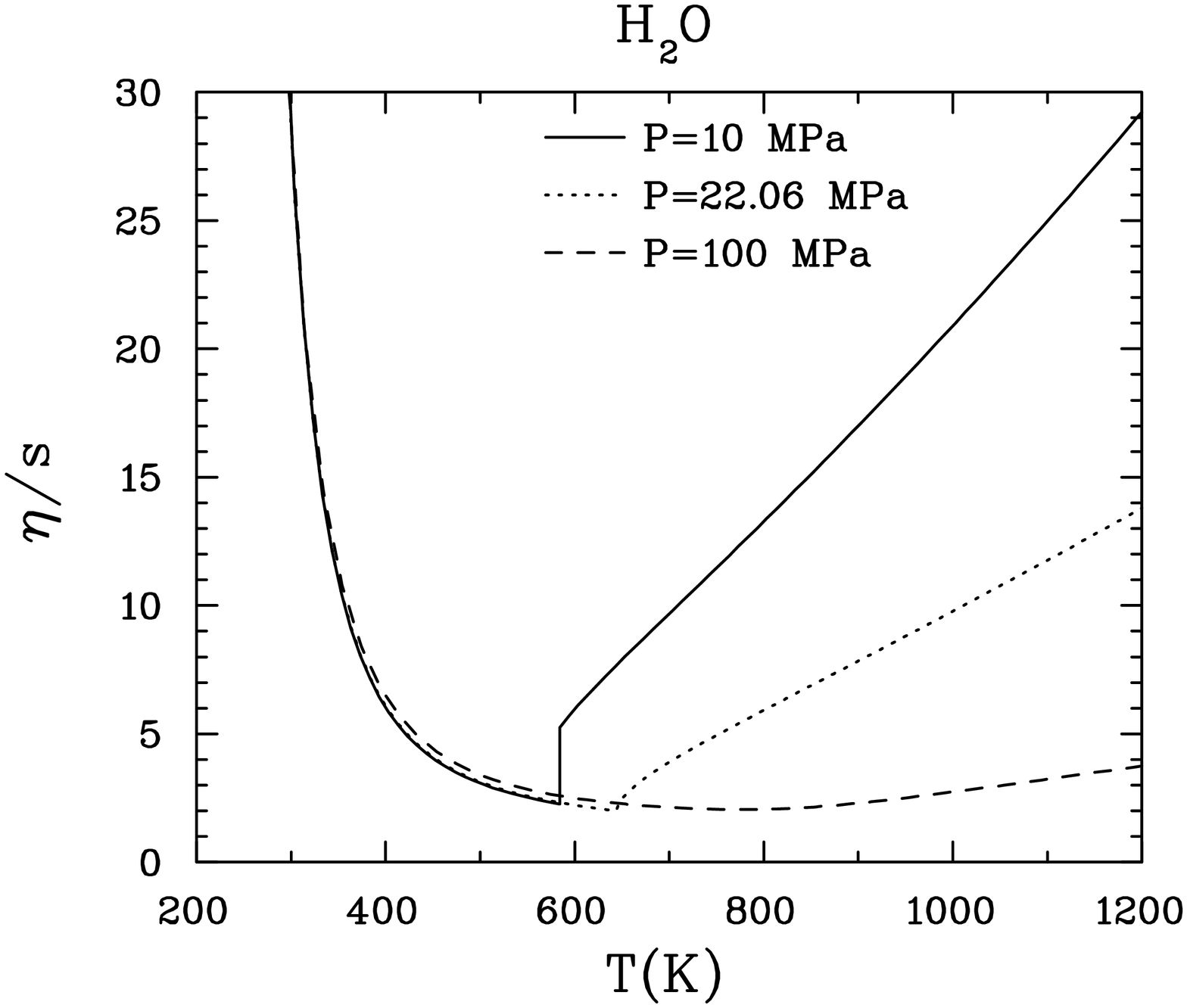}
 \caption{ The ratio $\eta/s$ as a function of $T$ for water with $s$ normalized 
such that $s(T=0)=0$.  The curves correspond to fixed pressures, one of them 
being the critical pressure, and the others being greater (100 MPa) and the 
other smaller (10 MPa).  Below the critical pressure there is a jump in the 
ratio, and above the critical pressure there is only a broad minimum. They were 
constructed using data from NIST and CODATA.  From Csernai, Kapusta and McLerran.}
 \label{figwater}
\end{figure}

\begin{figure}
 \centering
 \includegraphics[width=3.5in,angle=90]{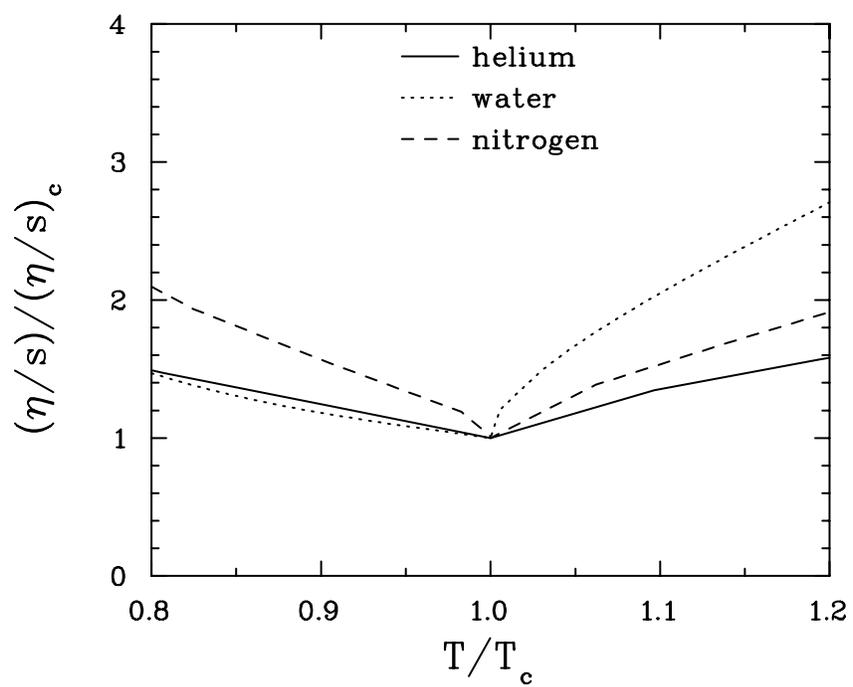}
 \caption{Scaled ratios along the critical isobar for helium, nitrogen and water obtained from the previous three figures.  There is no obvious universality.}
 \label{figscaled}
\end{figure}

\begin{figure}
 \centering
 \includegraphics[width=4.0in,angle=0]{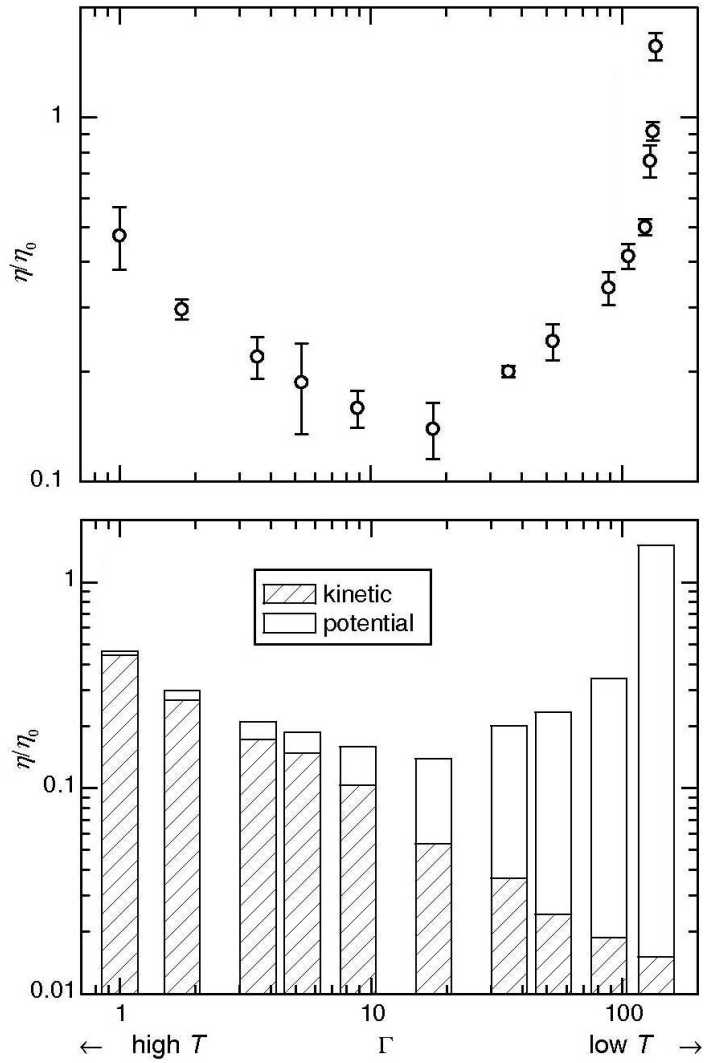}
 \caption{The shear viscosity versus the Coulomb coupling parameter for a 
two-dimensional Yukawa system in the liquid state.  The minimum in the viscosity occurs when it has approximately equal contributions from kinetic and potential components. From Liu and Goree.}
 \label{figGoree}
\end{figure}

\begin{figure}
 \centering
 \includegraphics[width=4.5in,angle=0]{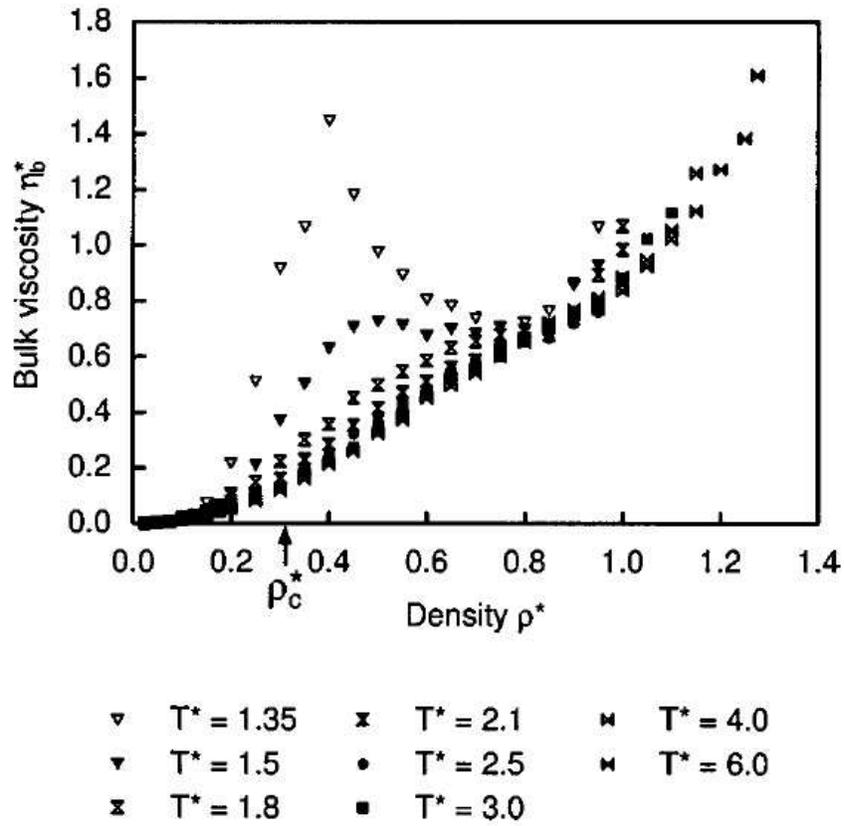}
 \caption{ The bulk viscosity versus density at fixed temperature for a classical liquid-gas phase transition using a Lennard-Jones potential.  The bulk viscosity has a peak near the critical point corresponding to the temperature $T^*=1.35$.  From Meier, Laesecke and Kabelac.}
 \label{figsuper}
\end{figure}

\begin{figure}
 \centering
 \includegraphics[width=3.5in,angle=90]{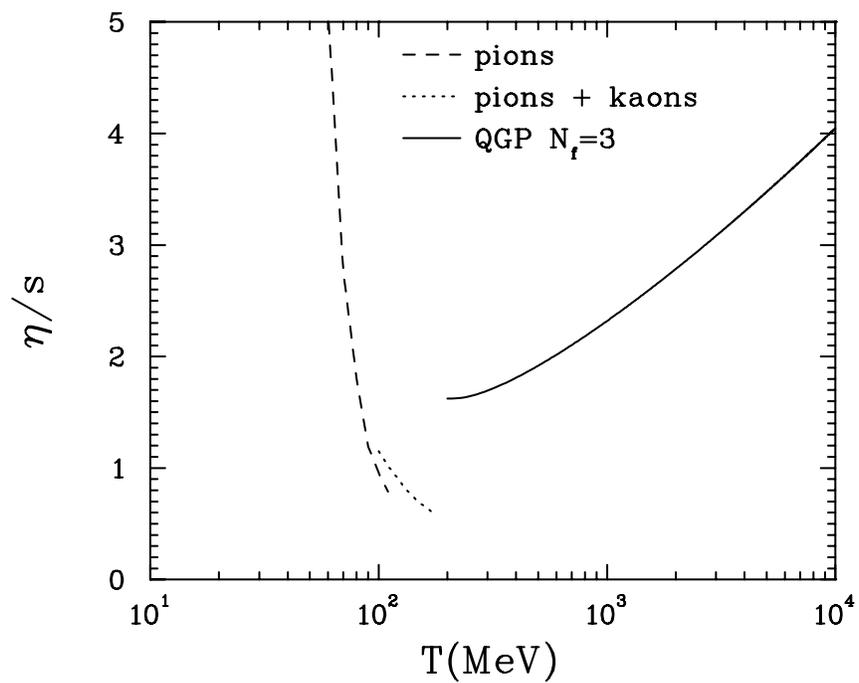}
 \caption{The ratio $\eta/s$ for the low temperature hadronic phase and for the 
high temperature quark-gluon phase.  Neither calculation is very reliable in the 
vicinity of the critical or rapid crossover temperature.  From Csernai, Kapusta and McLerran.}
 \label{figshearQCD}
\end{figure}

\begin{figure}
 \centering
 \includegraphics[width=3.5in,angle=90]{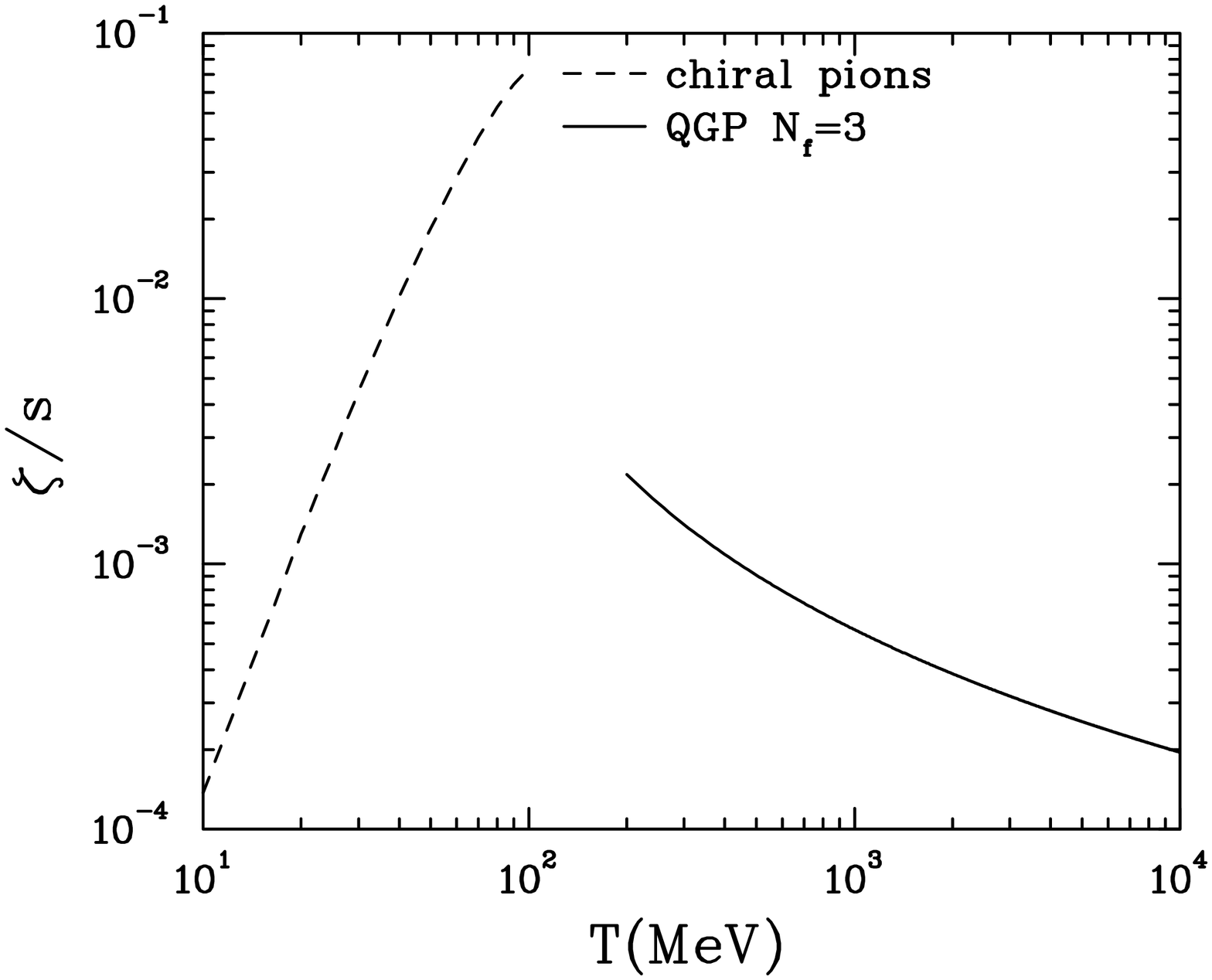}
 \caption{The ratio $\zeta/s$ for the low temperature hadronic phase represented by chiral pions and for the high temperature quark-gluon phase.  Neither calculation is very reliable in the vicinity of the critical or rapid crossover temperature.}
 \label{figbulkQCDv1}
\end{figure}

\begin{figure}
 \centering
 \includegraphics[width=3.5in,angle=90]{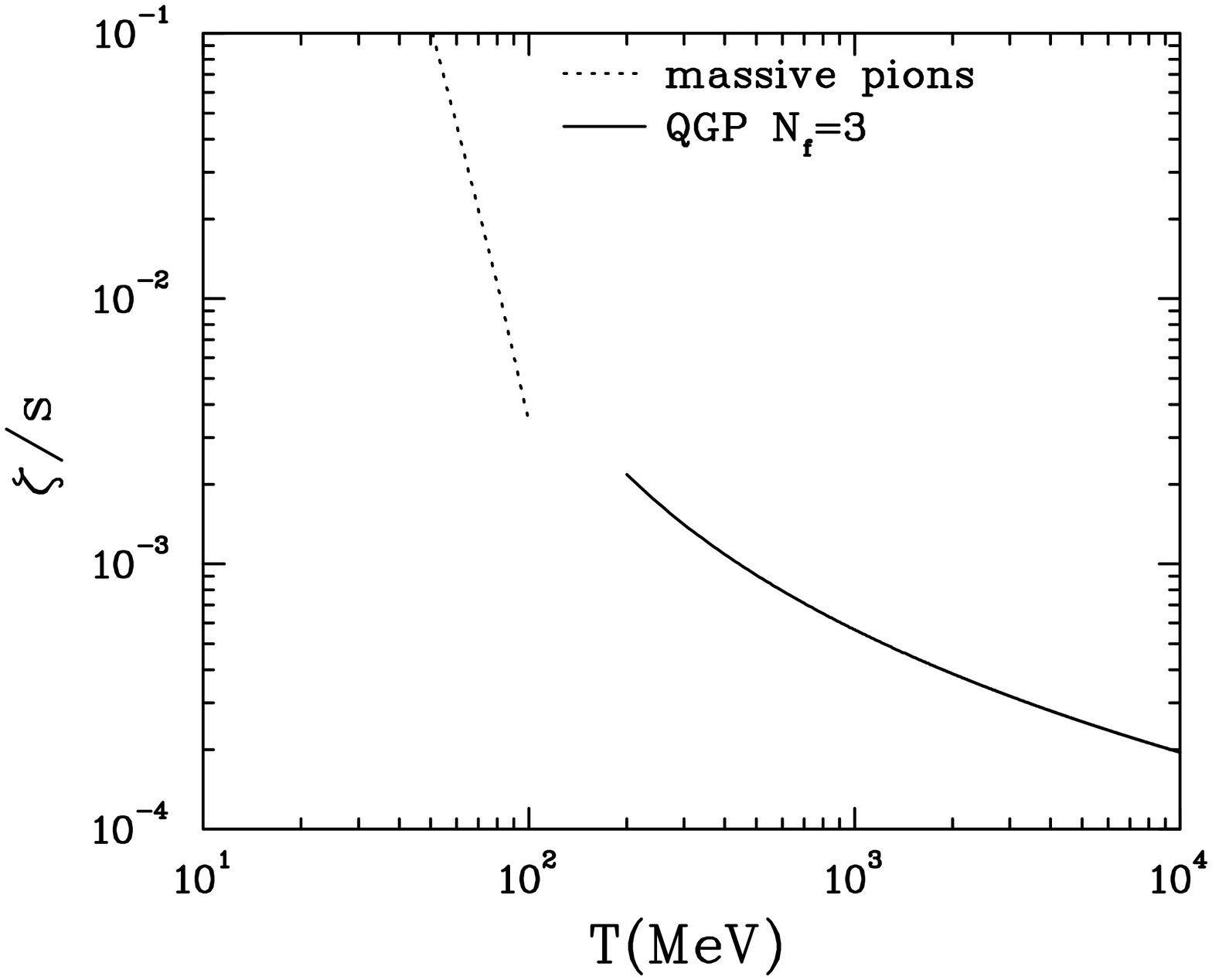}
 \caption{The ratio $\zeta/s$ for the low temperature hadronic phase represented by massive pions and for the high temperature quark-gluon phase.  Neither calculation is very reliable in the vicinity of the critical or rapid crossover temperature.}
 \label{figbulkQCDv2}
\end{figure}

\end{document}